\documentclass[final,authoryear,5p,times,twocolumn]{elsarticle}

\usepackage{epsfig}
\usepackage{graphicx}
\usepackage{subfig}
\usepackage{amsmath}
\usepackage{amssymb}
\usepackage{verbatim}
\usepackage{algorithm}
\usepackage{algorithmic}
\usepackage{booktabs}
\usepackage{multirow}
\usepackage{pbox} 
\usepackage{tabularx}
\usepackage{changepage}                 
\usepackage{lipsum}
\usepackage{tabularx}
\usepackage{hyperref}       
\usepackage{url}            
\usepackage{bm} 
\usepackage{color}
\newlength{\offsetpage}
	{\end{adjustwidth}}

\journal{Elsevier}

\begin{document}
\begin{frontmatter}



\title{Automatic Ischemic Stroke Lesion Segmentation from Computed Tomography Perfusion Images by Image Synthesis and Attention-Based Deep Neural Networks }

\author[address_1]{Guotai Wang\fnref{fn1}\corref{correspond_author}}
\fntext[fn1]{Equal Contribution}
\cortext[correspond_author]{Corresponding author}
\ead{guotai.wang@uestc.edu.cn}

\author[address_1,address_2]{Tao~Song\fnref{fn1}}
\author[address_3,address_4,address_5]{Qiang~Dong}
\author[address_3]{Mei~Cui}
\author[address_2]{Ning~Huang}
\author[address_1,address_2]{Shaoting~Zhang}

\address[address_1]{School of Mechanical and Electrical Engineering, University of Electronic Science and Technology of China, Chengdu, China}
\address[address_2]{SenseTime Research, Shanghai, China}
\address[address_3]{Department of Neurology, Huashan Hospital, Fudan University, Shanghai, China}
\address[address_4]{The State Key Laboratory of Medical Neurobiology, Fudan University, Shanghai, China}
\address[address_5]{Department of
Neurology, Jing’an District Centre Hospital of Shanghai, Shanghai, China}

\begin{abstract}
Ischemic stroke lesion segmentation from Computed Tomography Perfusion (CTP) images is important for accurate diagnosis of stroke in acute care units. However, it is challenged by low image contrast and resolution of the perfusion parameter maps, in addition to the complex appearance of the lesion. To deal with this problem, we propose a novel framework based on synthesized pseudo Diffusion-Weighted Imaging (DWI) from perfusion parameter maps to obtain better image quality for more accurate segmentation. Our framework consists of three components based on Convolutional Neural Networks (CNNs) and is trained end-to-end. First, a feature extractor is used to obtain both a low-level and high-level compact representation of the raw spatiotemporal Computed Tomography Angiography (CTA) images. Second, a pseudo DWI generator takes as input the concatenation of CTP perfusion parameter maps and our extracted features to obtain the synthesized pseudo DWI. To achieve better synthesis quality, we propose a hybrid loss function that pays more attention to lesion regions and encourages high-level contextual consistency. Finally, we segment the lesion region from the synthesized pseudo DWI, where the segmentation network is based on switchable normalization and channel calibration for better performance. Experimental results showed that our framework achieved the top performance on ISLES 2018 challenge and: 1) our method using synthesized pseudo DWI outperformed methods segmenting the lesion from perfusion parameter maps directly; 2) the feature extractor exploiting additional spatiotemporal CTA images led to better synthesized pseudo DWI quality and higher segmentation accuracy; and 3) the proposed loss functions and network structure improved the pseudo DWI synthesis and lesion segmentation performance. The proposed framework has a potential for improving diagnosis and treatment of the ischemic stroke where access to real DWI scanning is limited.

\end{abstract}


\end{frontmatter}


\section{Introduction}
Stroke is the most common cerebrovascular disease and one of the primary causes of mortality and long-term disability worldwide~\citep{Kissela2012}. Ischemic stroke is the most common type of stroke and  accounts for 75-85\% of all stroke cases, which is an obstruction of the cerebral blood supply and leads to tissue hypoxia (under-perfusion) and tissue death within few hours. The stages of stroke can be classified into acute (0 to 24h), sub-acute (24h to 2w) and chronic ($>$2w)~\citep{Gonzalez2011}. Early diagnosis and treatment in the acute stage is critical for recovery of the  stroke patient, and medical imaging is important for detection and quantitative assessment of stroke lesions, as well as eligible patient selection for thrombolysis or thrombectomy~\citep{Zaharchuk2012}.  

\begin{figure*}[t]
	\centering
	\includegraphics[width=1.0\linewidth]{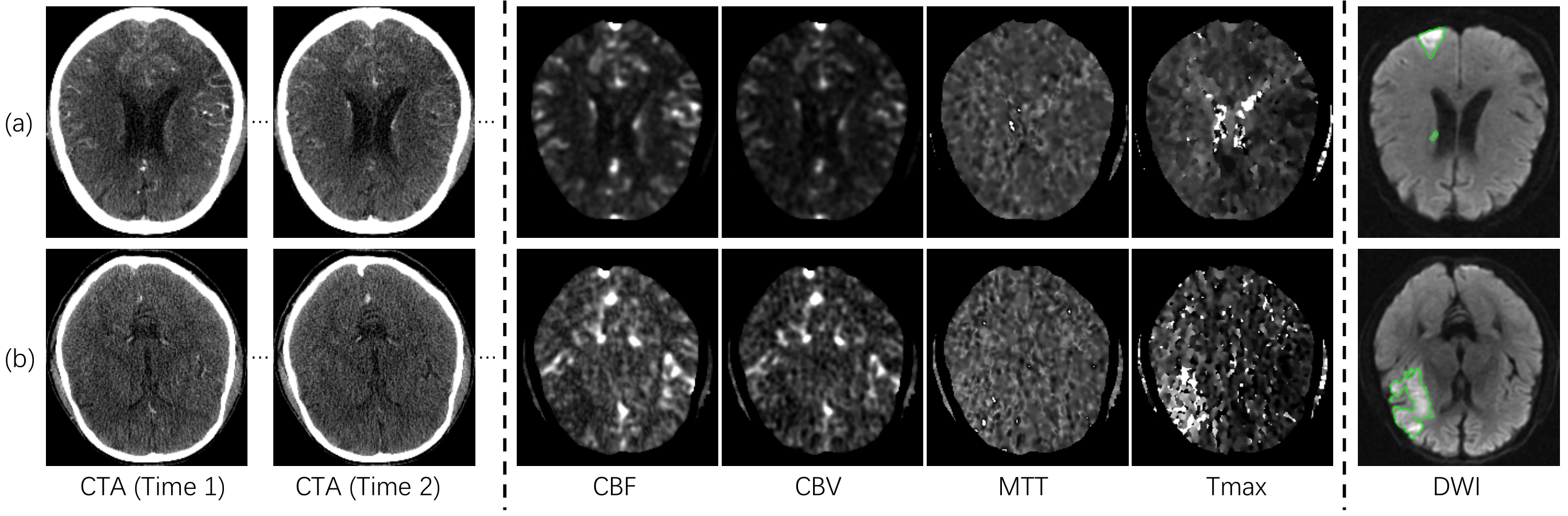}
	\caption{Examples of CTP and DWI images of two patients with ischemic stroke lesions. Column 1-2: CTA images at different time points during perfusion. Column 3-6: perfusion parameter maps.  Column 7: lesions delineated in DWI images. Note that we aim to segment the lesions from perfusion parameter maps, and DWI is not available at test time in our study.}
	\label{fig:example_images}
\end{figure*}

Among different medical imaging methods,
Magnetic Resonance Imaging (MRI) sequences such as Fluid-Attenuated Inversion Recovery (FLAIR), T1 weighted, T2 weighted,  and Diffusion-Weighted Imaging (DWI) are preferred imaging modalities for ischemic stroke lesions due to their good soft tissue contrasts. Especially, DWI is considered as the most sensitive method for detection of early acute stroke~\citep{Mezzapesa2006}. However, MR imaging including DWI is relatively slow and often not accessible for acute stroke patients. Alternatively, Computed Tomography Perfusion (CTP) imaging offers insights into cerebral hemodynamics and enables differentiation of salvageable penumbra from irrevocably damaged infarct core~\citep{Donahue2015}. CTP has advantages in speed and cost, leading to higher availability in acute care units~\citep{Gillebert2014}. In CTP imaging, a sequence of Computed Tomography Angiography (CTA) images (i.e., spatiotemporal 4D images) are acquired during the perfusion process, which results in perfusion parameter maps such as Cerebral Blood Flow (CBF), Cerebral Blood Volume (CBV), Mean Transit Time (MTT), Time to Peak (TTP, or Tmax) to help to identify ischemic stroke lesions. Examples of perfusion parameter maps of two ischemic stroke patients are shown in Fig~\ref{fig:example_images}.

Segmentation of stroke lesions from medical images can provide quantitative measurements of the lesion region, 
which is important for quantitative treatment decision procedures. Manual segmentation of the lesion is time-consuming with low inter-rater agreement, and automatic stroke lesion segmentation is more efficient and has a potential to provide more reliable and reproducible segmentation results~\citep{Maier2017}. 

Considering the limited speed and availability of MRI for acute stroke patients, we aim to segment ischemic stroke lesions automatically from CTP perfusion parameter maps, which has a potential for improving diagnosis and treatment of ischemic stroke in a timely fashion. However, this task is very difficult and the segmentation accuracy is confronted with a lot of challenges.  First, the appearance of stroke lesions varies considerably at different time, even within the same clinical stage of stroke~\citep{Gonzalez2011}. Second, the lesions have a large variation of location, shape, size and appearance in the brain, as shown in Fig.~\ref{fig:example_images}. Some lesions may be aligned with the vascular supply territories while others may not. The size of some small lesions can be only few millimeters, and some large lesions may cover a complete hemisphere~\citep{Maier2017}. The intensity is not homogeneous in the lesion region, and some other stroke-similar pathologies may lead to false positives in the segmentation result. Thirdly, compared with DWI, the perfusion parameter maps (CBF, CBV, MTT, and Tmax) are noisy with a lower spatial resolution, making it difficult to accurately identify the boundary of stroke lesions, as demonstrated in Fig.~\ref{fig:example_images}. In addition, the raw spatiotemporal 4D CTA images contain useful information of the ischemic stroke lesion but have a large data size. Using the perfusion parameter maps alone without considering the raw spatiotemporal CTA images may limit the segmentation accuracy, while directly taking raw spatiotemporal CTA images for lesion segmentation increases the computational cost. Therefore, extracting compact and useful features from the raw spatiotemporal CTA images is desirable for efficient and accurate ischemic stroke lesion segmentation.

Although automatic segmentation of ischemic stroke lesion has been widely studied, most of existing methods were proposed to deal with multi-modal MR images~\citep{Maier2017,Winzeck2018}. Only few works have been reported on ischemic stroke lesion segmentation from CTP images~\citep{Gillebert2014, Yahiaoui2016,Abulnaga2019}. Some old-fashion methods such as template-based methods~\citep{Gillebert2014} and fuzzy C-Means~\citep{Yahiaoui2016} are challenged by the complex appearance of stroke lesions. Recently, deep learning methods have achieved state-of-the-art performance for many medical image segmentation tasks~\citep{Shen2017}, and have been applied to ischemic stroke lesion segmentation from CTP images~\citep{Rittner2018,Abulnaga2019,Anand2018}. However, due to the above mentioned challenges, it remains difficult to segment the lesions directly from the perfusion parameter maps. 

Inspired by the fact that ischemic stroke lesions in DWI are easier to identify and segment than those in perfusion parameter maps, it is desirable to synthesize pseudo DWI images from perfusion parameter maps to help the segmentation task. Though a lot of methods have been proposed  for general medical image synthesis~\citep{Frangi2018}, synthesizing images with lesions is still not well addressed~\citep{Roy2010}, which is challenged by the complex variation of pathological lesions among patients. Especially, synthesizing pseudo DWI images from CTP images of ischemic stroke lesions has rarely been investigated.

This work is a substantial extension of our preliminary conference publication~\citep{TaoSong2018} that won the MICCAI 2018 ischemic stroke lesion segmentation (ISLES) challenge\footnote{\url{http://www.isles-challenge.org}}. In this paper, we provide detailed description and in-depth discussion of our segmentation framework and validate it with extensive experiments. The contribution of our work is summarized as follows.  

First, we propose a novel elaborated framework for automatic ischemic stroke lesion segmentation from CTP images based on synthesized pseudo DWI. Compared with using only CTP  perfusion parameter maps, our framework additionally exploits raw spatiotemporal CTA images for higher pseudo DWI synthesis quality and lesion segmentation accuracy.
Second, to make use of the raw spatiotemporal CTA images more efficiently, we propose a feature extractor that obtains more compact and high-level representation of the CTA images automatically, which helps to reduce the required memory and computational time and improve the performance of our segmentation method. 
Thirdly, we propose a novel method to synthesis pseudo DWI images with ischemic stroke lesions. We employ a high-level similarity loss function to encourage  the pseudo DWI to be close to the ground truth in terms of both local details and global context, and propose an attention-guided synthesis strategy so that the generator will focus more on the lesion part, which benefits the final segmentation. 
Last but not least, to segment lesions from our synthesized pseudo DWI, we propose a Convolutional Neural Network (CNN) with channel calibration and Switchable Normalization (SN)~\citep{Luo2019} that is suitable for small training batch size, and combine it with a novel attention-based and hardness-aware loss function that helps to obtain more accurate segmentation of ischemic stroke lesions. Experimental results show that our method achieved state-of-the-art performance on ISLES 2018 challenge and it outperformed direct segmentation from CTP perfusion parameter maps and contemporary image synthesis-based methods for ischemic stroke lesion segmentation from CTP images~\citep{Liu2018a}. 


\section{Related Works}
\subsection{Ischemic Stroke Lesion Segmentation}
Segmentation of ischemic stroke lesion from medical images has attracted increasing attentions in recent years~\citep{Rekik2012, Maier2017}, and most of them focus on segmentation from MR images. For example, the ISLES 2015-2017 challenges aimed at ischemic stroke lesion segmentation from multi-modal MR images including T1, T1-contrast, FLAIR and DWI sequences~\citep{Maier2017,Winzeck2018}. Some early works have used a range of methods for this segmentation task, such as Markov random field model~\citep{Kabir2007}, level set~\citep{Feng2016}, random forest~\citep{Mitra2014} and support vector machine~\citep{Maier2014}. However, their accuracy is challenged by the complicated segmentation problem~\citep{Maier2015}.
Recently, deep learning has been increasingly used for ischemic stroke lesion segmentation with better performance. For example, \cite{Kamnitsas2017} proposed a dual pathway 3D CNN combined with fully connected Conditional Random Field (CRF) for brain lesion segmentation. \cite{Cui2019} proposed an adapted mean teacher model to learn from a combination of annotated and unannotated MR images for the segmentation task.  \cite{Dolz2019} combined DWI and CTP  to segment ischemic stroke lesions and used a densely connected UNet with Inception modules~\citep{Szegedy2016} to handle the variation of lesion size. Despite their good performance, these methods rely on MRI and cannot be directly applied to stroke lesion segmentation from CTP images.  

There have been few works on the challenging task of segmentation of ischemic stroke lesion from CTA or CTP perfusion parameter maps~\citep{Rekik2012}.
 Some early works used histogram-based classifiers~\citep{Rekik2012} or template-based voxel-wise comparison ~\citep{Gillebert2014} to deal with this problem.  ~\cite{Yahiaoui2016} used a multi-scale contrast enhancement algorithm and fuzzy C-Means for this task. Recently, ~\cite{Abulnaga2019} used CNNs with pyramid pooling to combine global and local contextual information for this task, where a focal loss was employed to enable the CNNs to focus more on hard samples. 
However, due to the lower signal-to-noise ratio of CTP perfusion parameter maps compared with DWI, it remains challenging to automatically segment the ischemic stroke lesion from CTP images.

\subsection{Cross-Modality Medical Image Synthesis}
A range of works have investigated the problem of synthesizing medical images from another modality~\citep{Frangi2018}. For example,  ~\cite{Burgos2014} synthesized CT images from MRI through a multi-atlas information propagation scheme. ~\cite{Bahrami2016a} used dictionary learning to synthesis 7T-like images from 3T MRI. ~\cite{Jog2017} used regression random forest to synthesize T2 and FLAIR images from T1 images. Deep learning methods have also been increasingly used for medical image synthesis~\citep{Ker2017}, such as deep neural network-based synthesis methods~\citep{Nguyen2015} and deep adversarial learning-based approaches~\citep{Nie2018}. However, most of existing works deal with general cross-modality image synthesis and have not well investigated the more challenging problem of synthesizing medical images with pathological lesions. ~\cite{Roy2010} used an atlas-based method to synthesize FLAIR images with white matter lesions. ~\cite{Chartsias2017} proposed a CNN for synthesizing multi-modal MR images of brain lesions. The effectiveness of these methods for pseudo DWI synthesis from CTP perfusion parameter maps of stroke lesions has rarely been demonstrated.

\section{Method}
\begin{figure*}[t]
	\centering
	\includegraphics[width=1.0\linewidth]{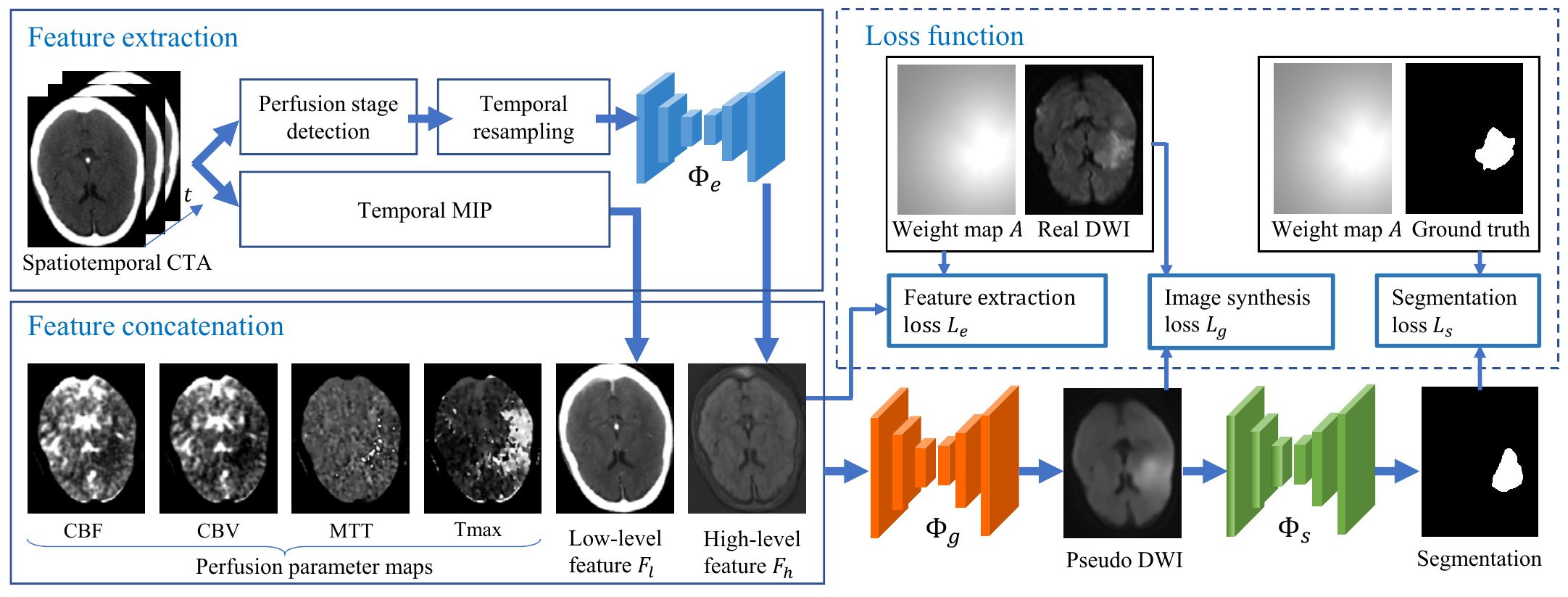}
	\caption{Illustration of the proposed framework for ischemic stroke lesion segmentation from CTP images. We extract additional low-level features based on temporal MIP and high-level features based on a CNN from raw spatiotemporal CTA images, and concatenate them with perfusion parameter maps. The concatenated images are used to generate pseudo DWI, from which the lesion is finally segmented. $\Phi_e$, $\Phi_g$ and $\Phi_s$ are three CNNs for high-level feature extraction, pseudo DWI generation and lesion segmentation, respectively. }
	\label{fig:framework}
\end{figure*}

The proposed framework for ischemic stroke lesion segmentation from CTP images is depicted in Fig.~\ref{fig:framework}. Due to the large inter-slice spacing (9.48~mm in average) of the experimental images, the proposed method operates on 2D slices. It consists of a feature extractor, a pseudo DWI generator and a final lesion segmenter. First, to efficiently deal with the large raw spatiotemporal CTA images and reduce the computational requirements, we design a high-level feature extractor that uses a CNN to obtain a compact representation of the raw spatiotemporal CTA images. Additionally, we make use of a temporal Maximal Intensity Projection (MIP) of the CTA images as a low-level feature.
Then, these features are concatenated with the perfusion parameter maps to serve as the input of the pseudo DWI generator, which obtains a pseudo DWI image with better contrast between the lesion and the background. To improve the synthesis quality near lesion regions, we use a high-level similarity-based loss function and enable the generator to pay more attention to the lesion. Finally, a segmenter takes the pseudo DWI image as input and produces a segmentation of the ischemic stroke lesion, where a CNN using channel calibration and switchable normalization trained with an attention-based and hardness-aware loss function is proposed to improve the performance. The three components are trained end-to-end. Details of these components will be described in the following.
\subsection{Feature Extraction from Raw Spatiotemporal CTA Images} \label{extractor} 
In CTP imaging, the raw spatiotemporal CTA images have been transformed into a simplified feature representation in terms of perfusion parameter maps including CBF, CBV, MTT and Tmax. Though these parameter maps are useful for detection of the stroke lesion, they are not a complete representation of the perfusion information in the raw spatiotemporal CTA images. Therefore, we do not ignore the raw spatiotemporal CTA images and try to mine some additional features that are useful in the segmentation task.  

Let $I(x,y,z,t)$ represent a raw spatiotemporal CTA image obtained during the perfusion, where $t\in [0, 1, 2, ..., T-1]$ and $T$ is the total number of time points. 
Considering that the raw spatiotemporal CTA image has a large data size due to a large value of $T$, we use a feature extractor to obtain an additional low-level feature and a compact and high-level representation of the raw spatiotemporal CTA image to make an efficient use of it. 
The feature extraction method is shown in Fig.~\ref{fig:framework}. We extract both a  manually designed low-level feature and a high-level feature that is automatically learned by a CNN.

First, the maximal intensity value of a voxel during perfusion may contain information related to the ischemic stroke lesion~\citep{Murayama2018}. Therefore, in addition to the standard perfusion parameter maps, we apply a Maximal Intensity Projection (MIP) along the temporal axis to $I$ to obtain a low-level feature map $F_l$:
\begin{align}\label{eq:f_l}
F_l = \max_{t}~I(x,y,z,t)
\end{align}
\begin{figure}[t]
	\centering
	\includegraphics[width=1.0\linewidth]{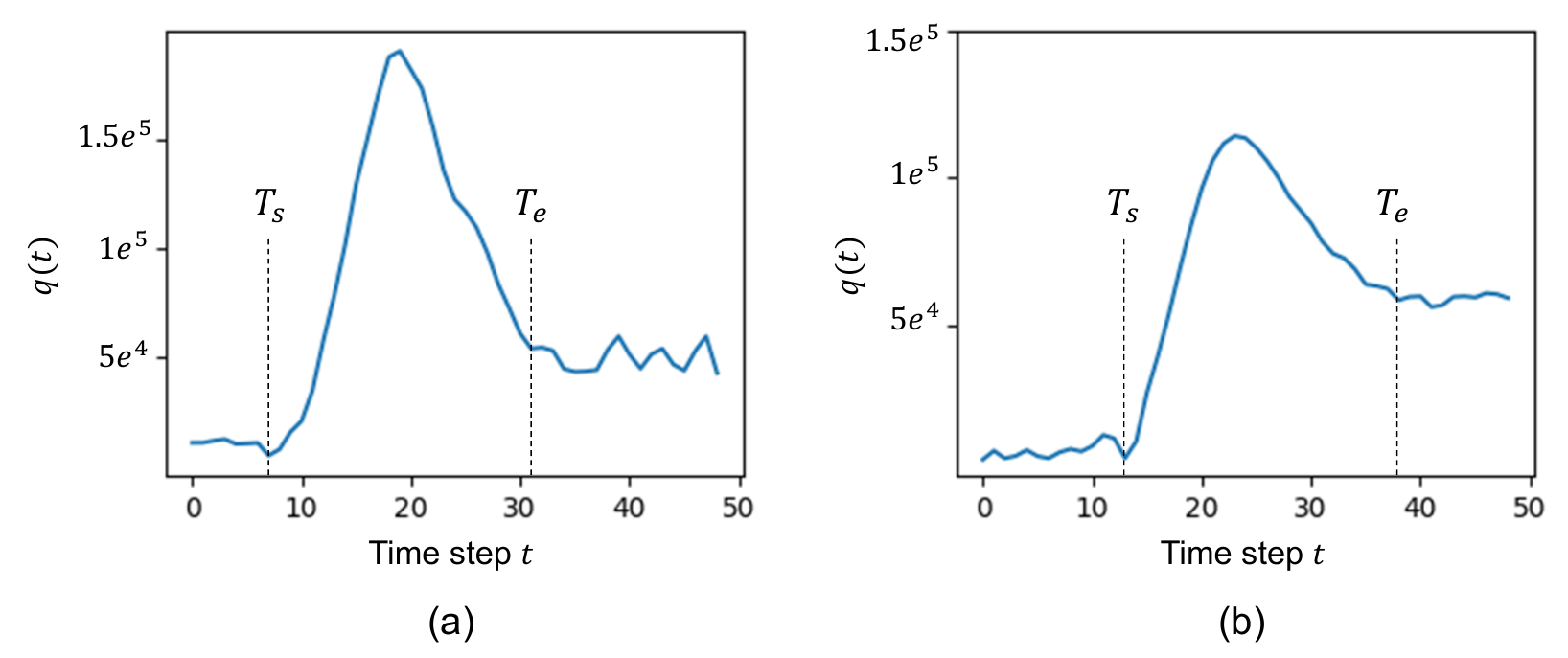}
	\caption{Illustration of start time ($T_s$) and end time ($T_e$) detection of the perfusion stage. }
	\label{fig:ts_te}
\end{figure}

Second, we use a CNN to extract high-level features of the raw spatiotemporal CTA image due to CNNs' good performance in automatic feature extraction~\citep{Shen2017}. Though the start and end time points of  perfusion do not affect the MIP image in theory, they are important for the high-level feature extractor, as the CNN is designed to take the frames during the perfusion as input. To reject frames that are not perfused in the raw spatiotemporal CTA image, we need first to detect these two time points.
We define a curve of accumulated intensity over time as $q(t) = \sum_{x,y,z}I(x,y,z,t)$. Let $T_s$ and $T_e$ denote the estimated start and end time points of the perfusion respectively. They are determined by the following rules:
\begin{align}
	T_s = \min~\Big\{~t~|~0~\le t < T - K, \sum_{k = 0}^{K-1} \mathcal{H}\big(q'(t + k)\big) =K \Big\}
\end{align}
\begin{align}
T_e = \max~\Big\{~t~|~K \le t < T, \sum_{k = 0}^{K-1} \mathcal{H}\big(q'(t - k)\big) =0 \Big\}
\end{align}
where $\mathcal{H}(\cdot)$ is the Heaviside function that obtains 0 for negative inputs and 1 for positive inputs. $q'(t)$ is the first derivative of $q(t)$, and $K$ is a positive integer value which is 5 in this paper. Therefore, $T_s$ is defined as the earliest time point where the first derivative of $q(t)$ keeps positive for its following $K$ consecutive  time points, and $T_e$ is defined as the latest time point where the first derivative of $q(t)$ keeps negative for its preceding $K$ consecutive  time points. Fig.~\ref{fig:ts_te} shows the curve of $q(t)$ with $T_s$ and $T_e$ in two cases.

We extract the frames between $T_s$ and $T_e$ and obtain a temporally cropped subsequence that corresponds to the perfusion stage of the raw spatiotemporal CTA image. 
As the duration of the perfusion stage has a variation among different subjects, the temporally cropped subsequence can have different time point numbers along the temporal axis. To deal with this problem and to reduce the computational cost, we uniformly down-sample the temporally cropped subsequence along the temporal axis into a fixed time point number of $C_e$. The temporally cropped and down-sampled CTA image is referred to as $I^*$, which is used as the input of a CNN for high-level feature extraction. 

Let $C_e\times D \times H \times W$ represent the size of $I^*$, where $D$, $H$ and $W$ represent the spatial depth, height and width of the input 4D image $I^*$ respectively. 
We treat $I^*$ as a multi-channel 3D volume and  use a 2D CNN for high-level feature extraction from each slice, as the images have a large inter-slice spacing (9.48~mm in average) in this study. Specifically, we used the UNet~\citep{Ronneberger2015} for the high-level feature extraction due to its good performance in a range of  tasks~\citep{Abdulkadir2016,Li2017c,Isensee2018}. The UNet consists of an encoding path and a decoding path. The encoding path uses convolution and down-sampling through max-pooling layers to obtain features at different scales with reduced spatial resolution, and the decoding path uses up-sampling (deconvolution) layers to recover the spatial resolutions. We set the output channel of the extractor CNN to 1. Let $F_h$ denote the CNN's output and it has a size of $1 \times D \times H \times W$, which  is a high-level representation of the input spatiotemporal CTA image $I^*$.
\begin{align}\label{eq:f_h}
F_h = \Phi_e(I^*, \theta_e)
\end{align}
where $\Phi_e$ represents the feature extraction network and $\theta_e$ denotes the set of parameters of the network.
 
\subsection{Pseudo DWI Synthesis from CTP Images} \label{synthesis}

Inspired by recent works on CNN-based image synthesis with state-of-the-art performance~\citep{Frangi2018}, we also use CNNs to generate pseudo DWI images, and select UNet~\citep{Ronneberger2015} as the backbone network structure due to its good performance.
Differently from previous works that synthesized pseudo DWI images only from  CTP perfusion parameter maps including CBF, CBV, MTT and TMax~\citep{Liu2018a}, we additionally take advantage of the extracted low-level and high-level features ($F_l$ and $F_h$) so that more information from the raw spatiotemporal CTA image can help to improve the quality of the synthesized pseudo DWI. Let $F_o$ represent the concatenation of CBF, CBV, MTT and TMax. The input of our generator is a concatenation of $F_o$, $F_l$ and $F_h$ and thus it has six channels. The generated pseudo DWI can be represented as:
\begin{align}
I_g = \Phi_g(F_o, F_l, F_h, \theta_g)
\end{align}
where $\Phi_g$ represents the pseudo DWI generation network and $\theta_g$ denotes its parameter set.

\begin{figure}[t]
	\centering
	\includegraphics[width=1.0\linewidth]{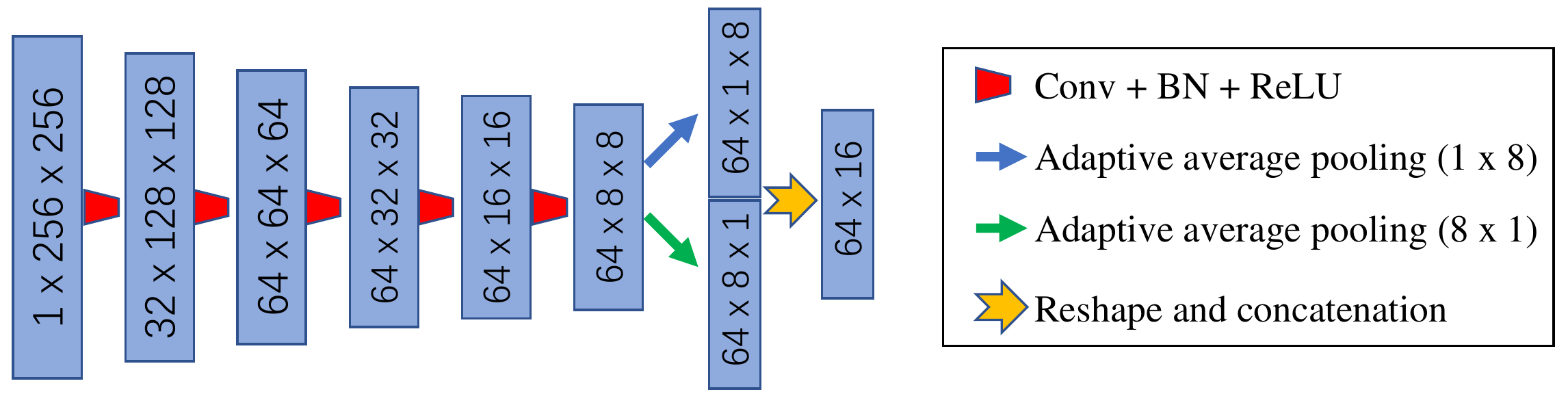}
	\caption{Structure of the encoder $\Phi_c$ to obtain a high-level representation of an input image. The convolution kernels have a size of 3$\times$3 and a stride of 2$\times$2.  }
	\label{fig:encoder}
\end{figure}

Let $I_d$ represent the DWI ground truth for synthesis. To train the generator $\Phi_g$ so that it can focus on the lesion region and the output $I_g$ has a high-level similarity to the ground truth $I_d$, we propose a novel loss function $L_g(I_g, I_d)$ that combines a low-level weighted pixel-wise loss $\ell_l(I_g, I_d)$ and a high-level contextual loss $\ell_h(I_g, I_d)$:
\begin{align} \label{eq:generator_loss}
L_g(I_g, I_d) = \ell_l(I_g, I_d) + \gamma \ell_h(I_g, I_d) 
\end{align}
\begin{align} \label{eq:generator_loss_a}
\ell_l(I_g, I_d) = ||A\cdot(I_g - I_d)||_2 
\end{align}
\begin{align} \label{eq:generator_loss_b}
\ell_h(I_g, I_d) = || \Phi_c(I_g, \theta_c) - \Phi_c(I_d,\theta_c)||_1
\end{align}
where $\gamma$ is a weighting parameter for the contextual loss and $A$ is a spatial weight map. $||\cdot||_2$ and $||\cdot||_1$ are the $L2$-norm and $L1$-norm respectively.
As we follow the common practice of using the Peak Signal-to-Noise Ratio (PSNR) that is related to Mean Square Error (MSE) as one of the metrics to evaluate the image quality, here $L2$-norm  is used for pixel-level loss so that minimizing the $L2$-norm corresponds to maximizing the PSNR. On the other hand, as $L1$-norm treats each element equally while $L2$-norm assigns higher weights (i.e., by squaring) to larger prediction errors that may be caused by outliers,  $L1$-norm has a higher robustness than $L2$-norm~\citep{Ghosh2017}. Therefore, we use $L1$-norm for the high-level contextual loss.
$\Phi_c$ is a CNN-based encoder with a parameter set $\theta_c$ and it converts $I_g$ and $I_d$ into their  high-level and compact (i.e., low dimensional) representations, respectively. As $\ell_l(\cdot)$ operates on individual voxel-wise predictions and does not guarantee global and high-level consistency, $\ell_h(\cdot)$ based on the encoder $\Phi_c$ helps to overcome this problem by encouraging closeness between the lower dimensional non-linear projections of $I_g$ and $I_d$. 
Our encoder $\Phi_c$ consists of five convolutional layers and two adaptive average pooling layers, and its output is a vector of length 16. Details of $\Phi_c$ are shown in Fig.~\ref{fig:encoder}. 

As our final goal is to segment the ischemic stroke lesion, a good synthesis quality around the lesion region is desirable. Therefore, we use the voxel-wise weight map $A$ to make the generator pay more attention to the lesion region and less attention to the background. Let $\mathcal{F}$ denote the set of lesion foreground voxels, and $Eud(i,\mathcal{F})$ denote the shortest Euclidean distance between a voxel $i$ and $\mathcal{F}$. We use $A_i$ to represent the weight of voxel $i$ in the weight map $A$:
\begin{align} \label{eq:weight}
A_i = \begin{cases}
w, & \text{if } i\in \mathcal{F}\\
0.5 + \frac{exp(-Eud(i,\mathcal{F})/D)}{exp(-Eud(i,\mathcal{F})/D) + 1},              & \text{otherwise}
\end{cases}
\end{align}
where $w \ge 1$ is the weight for foreground voxels and $D$ is a positive parameter that controls the sharpness of the weight for background voxels. $A_i$ decays gradually with the increase of $Eud(i,\mathcal{F})$, i.e., the weights for voxels that are further from the lesion region are lower. An example of $A$ is shown in Fig.~\ref{fig:framework}.

\subsection{SLNet: Stroke Lesion Segmentation Network with Switchable Normalization and Channel Calibration} \label{segenter}
Our segmentation network takes the synthesized pseudo DWI image $I_g$ as input and outputs a binary segmentation of the ischemic stroke lesion. Let $\Phi_s$ represent the segmentation network and $\theta_s$ denote its parameter set. The segmentation network's output probability map is formatted as:
\begin{align} 
P = \Phi_s(I_g, \theta_s)
\end{align}
where $P$ has $C$ channels and $C$ equals to the class number, which is 2 in our binary segmentation task. We select the UNet structure~\citep{Ronneberger2015} as the backbone and extend it in two aspects to obtain a better performance.
\begin{figure}[t]
	\centering
	\includegraphics[width=1.0\linewidth]{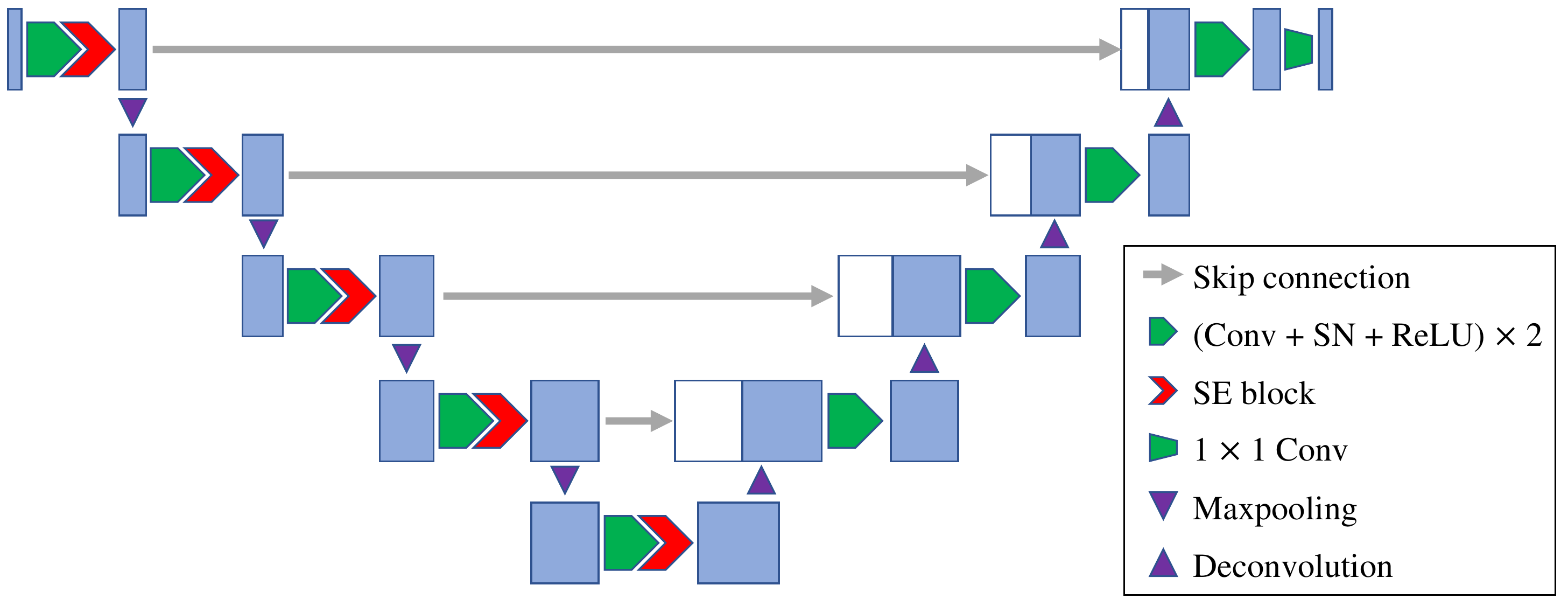}
	\caption{The proposed SLNet for ischemic stroke lesion segmentation with Switchable Normalization (SN) and Squeeze-and-Excitation (SE) blocks.  }
	\label{fig:slnet}
\end{figure}

First, we replace Batch Normalization (BN) layers with switchable normalization~\citep{Luo2019} layers, which learn to automatically select suitable normalizers for different normalization layers of a CNN. Compared with traditional batch normalization, switchable normalization is more robust to a wide range of batch sizes and more suitable for small batch sizes~\citep{Luo2019}. In our segmentation task, the large input patches and dense feature maps take a lot of memory, which limits the batch size to a small number. Therefore, switchable normalization is preferred to batch normalization.  
Second, as different channels in a feature map may have different importance, we use a Squeeze-and-Excitation (SE) block~\citep{Hu2017b} based on channel attention to calibrate channel-wise feature responses. The SE block explicitly models inter-channel dependencies by learning an attention weight for each channel so that the network relies more on the most important channels for segmentation. We use an SE block after each convolution block  in the encoding path of the UNet~\citep{Ronneberger2015}. The proposed network is referred to as SLNet, which is shown in Fig.~\ref{fig:slnet}.

To deal with the large range of the ischemic stroke lesion size and challenging training samples for the segmentation task, we propose a novel hybrid loss function to train the segmentation network. Let $Y$ denote the one-hot ground truth label with channel number $C$. We use $P^c_i$ and $Y^c_i$ to denote the probability of voxel $i$ belonging to class $c$ in the prediction output and the ground truth respectively. The proposed loss function is a combination of a weighted cross entropy loss function $L_{WCE}$ and a hardness-aware generalized Dice loss function $L_{HGD}$:
\begin{align} \label{eq:segment_loss}
L_s(P, Y) = L_{WCE}(P, Y, A) + L_{HGD}(P, Y) 
\end{align}
\begin{align}\label{eq:weighted_ce}
L_{WCE}(P, Y, A) = \frac{\sum_i^N A_i \Big(\sum_c^C -Y_i^c\log P_i^c\Big)}{\sum_i^N A_i}
\end{align}
\begin{align}\label{eq:hgd}
L_{HGD}(P,Y)  = - \log \Big(1 - L_{GD}(P, Y)\Big)
\end{align}
\begin{align}\label{eq:gd}
L_{GD}(P, Y) = 1 - 2\frac{\sum_c^C m_c \sum_i^N Y_i^c P_i^c}
{\sum_c^C m_c \sum_i^N (Y_i^c + P_i^c)}
\end{align}
where $N$ is the number of voxels. $A$ is a voxel-wise weight map, and we use the same one as defined in Eq.~\ref{eq:weight}, which drives the segmentation network to pay more attention to the lesion region than the background. $L_{GD}$ is the generalized Dice loss that automatically balances different classes by defining a class-wise weight $m_c = 1 / (\sum_i^N Y_i^c)^2$~\citep{Sudre2017}. Inspired by the focal loss~\citep{Lin2017} that automatically penalizes hard samples in object detection tasks, we use $-\text{log} (1-L_{GD})$ in Eq.~\ref{eq:segment_loss} that has the same monotonicity as $L_{GD}$ but gets higher gradient values for large $L_{GD}$ values, so that our segmentation loss function is also aware of hard image samples.

\subsection{End-to-End Training } \label{total_loss}
The overall pipeline of our feature extractor $\Phi_e$, pseudo DWI generator $\Phi_g$,  image context encoder $\Phi_c$ and the final segmentation network $\Phi_s$ can be jointly trained in an end-to-end fashion. The overall loss function for training is therefore defined as: 
\begin{align} \label{eq:overall_loss}
L = L_s(P, Y) + \alpha L_g(I_g, I_d) + \beta L_e(F_h, I_d)
\end{align}
where $\alpha$ and $\beta$ are weighting parameters. The segmentation loss function $L_s(P, Y)$ is defined in Eq.~\ref{eq:segment_loss}
and the pseudo DWI synthesis loss function $L_g(I_g, I_d)$ is defined in Eq.~\ref{eq:generator_loss}. To obtain better synthesized pseudo DWI and lesion segmentation results, we add an extra explicit supervision on  $F_h$ that is the output of the feature extractor $\Phi_e$. Therefore, we introduce a loss $L_e(F_h, I_d) = L_g(F_h, I_d)$  to encourage the similarity between $F_h$ and $I_d$. The end-to-end training will update $\theta_e$, $\theta_g$, $\theta_c$ and $\theta_s$ simultaneously.

\section{Experiments and Results}
\label{headings}

\subsection{Data and Implementation}

We used the dataset from ISLES challenge 2018\footnote{\url{http://www.isles-challenge.org}} to validate our segmentation framework. The ISLES 2018 dataset includes CTP scanning of 103 patients in two centers who presented within 8 hours of stroke onset. For the CTP scanning, a contrast agent was administered to the patient and then sequential CTA images were acquired 1-2 seconds apart. Then the perfusion parameter maps  CBF, CBV, MTT and Tmax were derived from the raw spatiotemporal CTA images. An MRI DWI scanning was obtained within 3 hours after the CTP scanning for each patient. The intra-slice pixel spacing ranged from 0.80~mm $\times$ 0.80~mm to 1.04~mm $\times$ 1.04~mm, with a slice size of 256 $\times$ 256. 
The inter-slice spacing ranged from 4.0~mm to 12.0~mm with a mean value of 9.48~mm. 
The slice number ranged from 2 to 22 with a mean value of 5.34, and the time point number for CTA ranged from 43 to 64 with a mean value of 47.18. For high-level feature extraction, all the CTA images were temporally cropped and down-sampled with an output time point number of $C_e$ = 6. For preprocessing, intensity values in each DWI volume were scaled to (0, 1) based on the minimal value and the 99-th percentile.
Manual delineation of the stroke lesion from DWI images given by an expert was used as the segmentation ground truth. The training set consisted of 94 scannings of CTP and DWI from 63 patients.
The testing set consisted of 62 CTP scannings from 40 patients, for which DWI images were not provided to participants of the challenge.

Our segmentation framework was implemented by PyTorch\footnote{\url{https://pytorch.org}} with an NVIDIA TITAN X GPU with 12 GB memory. The weights of all networks were initialized by Xavier method~\citep{Glorot2010} and trained with the RMSprop optimizer~\citep{Tieleman2012}, a batch size of 5 and 300 epochs.  We initialized the learning rate as 0.002 and reduced it by a factor of 0.2 after 180 epochs. The parameter setting was: $\alpha = 1.0$, $\beta = 1.0$, $\gamma = 1.2$, $w = 1.5$ and $D$ = 50.

To quantitatively evaluate the quality of the generated pseudo DWI images, we measured the Structure Similarity (SSIM) and Peak Signal-to-Noise Ratio (PSNR) between the DWI ground truth and the generated pseudo DWI. These two metrics were calculated both globally (i.e., in the entire image region) and locally (i.e., in the region around the ground truth lesion). The local SSIM and PSNR  are helpful for the assessment of our method's ability to generate high-quality lesion regions in a pseudo DWI image. 

For quantitative evaluations of the segmentation accuracy, we use Precision, Recall, Dice score, Hausdorff Distance (HD) and Average Symmetric Surface Distance (ASSD). 
\begin{align}\label{eq:dice}
	Dice = \frac{2\times TP}{
	2\times TP + FN + FP}
\end{align}
where $TP$, $FP$ and $FN$ are true positive, false positive and false negative respectively. 
\begin{align}\label{eq:hausdorff}
HD = \max{\bigg\{\max_{s\in S}d(s,G), \max_{g\in G}d(g,S) \bigg\}}
\end{align}
\begin{align}\label{eq:assd}
ASSD = \frac{1}{|S| + |G|}{\Bigg(\sum_{s\in S}d(s,G) + \sum_{g\in G}d(g,S) \Bigg)}
\end{align}
where $S$ and $G$ denote the set of surface points of a segmentation result and the ground truth respectively. $d(s,G)$ is the shortest Euclidean distance between a point $s\in S$ and all the points in $G$. 

\subsection{Ablation Studies}
\label{ablation_study}
We first conducted ablation studies to validate different components of our segmentation framework. Since the ground truth segmentations of ISLES testing images were not available for participants, we split the official ISLES training set at patient level into our local training, validation and testing sets, which contained images from 65, 6 and 23 scannings respectively. In this section, we report the experimental results obtained from our local testing images.

\subsubsection{Comparison of Different Loss Functions for Pseudo DWI Synthesis}
\begin{figure*}[t]
	\centering
	\includegraphics[width=0.8\linewidth]{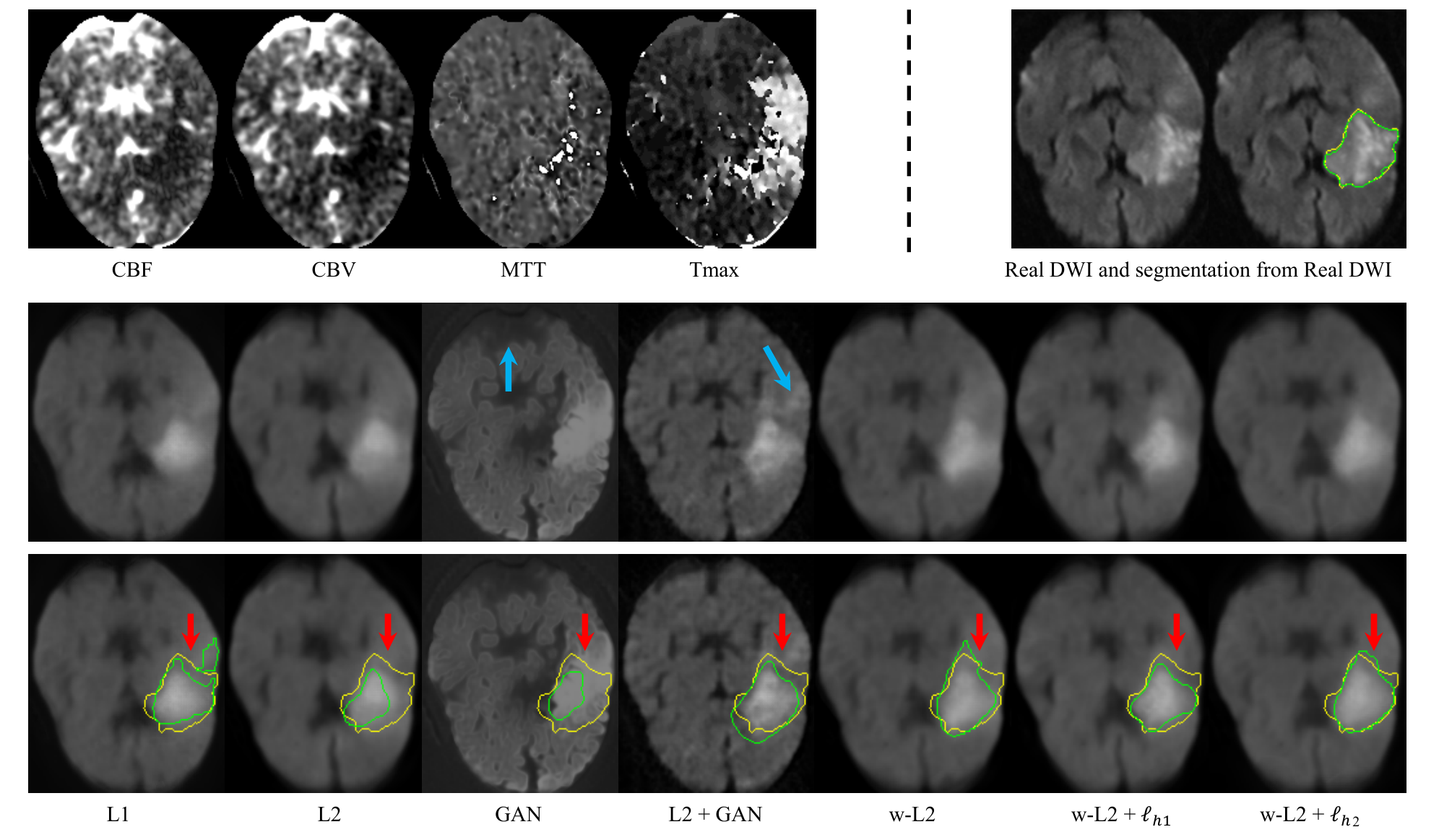}
	\caption{Visual comparison of pseudo DWI synthesis result (the second row) obtained by different loss functions and their effect on segmentation (the third row). First row: Concatenation of perfusion parameter maps ($F_o$) was used as the input of UNet for synthesis.  w-L2: Weighted L2 loss defined in Eq.~\ref{eq:generator_loss_a}. w-L2 + $\ell_{h1}$: Proposed hybrid loss based on Eq.~\ref{eq:generator_loss} and Eq.~\ref{eq:generator_loss_b}.  Light blue arrows highlight artifacts obtained by GAN-based methods, and red arrow highlight the segmentation differences. Green and yellow curves show segmentation result and the ground truth, respectively. }
	\label{fig:synthesis_loss_visual1}
\end{figure*}
First, we investigated the effect of different loss functions on pseudo DWI synthesis from perfusion parameter maps $F_o$, i.e., concatenation of CBF, CBV, MTT and Tmax. The proposed loss function $Ig$ (Eq.~\ref{eq:generator_loss}) based on weighted L2 loss and high-level contextual loss (Eq.~\ref{eq:generator_loss_b}) is referred to as w-L2 + $\ell_{h1}$, which is compared with 1) L1 loss that refers to $\ell_l$ in  Eq.~\ref{eq:generator_loss_a} being defined as L1 norm with $A_i = 1$ for every voxel; 2) L2 loss as defined in Eq.~\ref{eq:generator_loss_a} with $A_i = 1$ for every voxel; 3) w-L2 loss that refers to Eq.~\ref{eq:generator_loss_a} with weight coefficients defined in Eq.~\ref{eq:weight}; 4) adversarial training with Generative Adversarial  Networks (GAN), which is referred to as GAN; 5) L2 + GAN that combines L2 loss and GAN loss and 6) w-L2 + $\ell_{h2}$ that refers to a variant of the proposed $I_g$ with $\ell_h$ based on L2 norm. For the GAN method, we used the LSGAN framework proposed by ~\cite{Mao2016a}, and used a multi-scale discriminator~\citep{Ting-ChunWang2018} to guide the generator (i.e. UNet) to produce realistic local details and global appearance.

\begin{table*}[t]
	\centering
	\scriptsize 
	\caption{Quantitative evaluation of different training loss functions for pseudo DWI synthesis and their effect on segmentation. Concatenation of the CTP perfusion parameter maps ($F_o$) was used as the input for synthesis. }
	\begin{tabular}{l|c|c|c|c|c}
		\hline
		Loss & Global SSIM & Local SSIM & Global PSNR & Local PSNR  & Dice (\%) \\  \hline
		L1  & 0.82$\pm$0.11 & 0.47 $\pm$0.16 & 19.36$\pm$4.11  & 13.60$\pm$4.25 & 49.45$\pm$21.20 \\
		L2  & 0.83$\pm$0.11 & 0.51$\pm$0.17 & 19.41$\pm$3.63  & 13.82$\pm$4.18 & 50.04$\pm$19.38 \\
		GAN   & 0.81$\pm$0.11 & 0.37$\pm$0.17 & 17.57$\pm$4.27 & 13.45$\pm$4.75 & 41.53$\pm$25.08 \\
		L2 + GAN  & 0.78$\pm$0.12 & 0.52$\pm$0.14 & 18.30$\pm$3.91 & 13.22$\pm$4.52 & 48.77$\pm$20.73\\		
		w-L2& \textbf{0.83$\pm$0.09}  & 0.53$\pm$0.15  & \textbf{19.43$\pm$3.42} & \textbf{13.99$\pm$4.01}  & 50.95$\pm$21.03 \\
		w-L2  + $\ell_{h1}$  & 0.83$\pm$0.10 & \textbf{0.54$\pm$0.13} & 19.26$\pm$3.32  & 13.20$\pm$3.38  & \textbf{51.25$\pm$17.43} \\
		w-L2  + $\ell_{h2}$  & 0.83$\pm$0.11 &  0.53$\pm$0.15 & 19.22$\pm$3.40  & 13.80$\pm$3.41  & 51.02$\pm$21.41\\	\hline
	\end{tabular}
	\label{tab:synthesis_loss}.
\end{table*}

Fig.~\ref{fig:synthesis_loss_visual1} shows a visual comparison of pseudo DWI generated by UNet trained with different loss functions, where the input images were perfusion parameter maps ($F_o$) for these variants. 
The synthesized pseudo DWI images are shown in the second row. It can be observed that $L1$ and $L2$ obtained similar results with ambiguous lesion boundary. The use of w-L2 and w-L2 + $\ell_{h1}$ losses helps to obtain clearer lesion boundary respectively. The results of GAN and L2 + GAN are less smoothed, but include some large artifacts as highlighted by the light blue arrows.
We additionally investigated the effect of the synthesized pseudo DWI images on segmentation,  where we used the standard cross entropy loss to train a segmentation model (i.e., UNet~\citep{Ronneberger2015}) with each type of these synthesized pseudo DWI images respectively.
 The last row in  Fig.~\ref{fig:synthesis_loss_visual1} shows that the segmentation based on synthesized pseudo DWI images obtained by w-L2 + $\ell_{h1}$ is more accurate than the others, as highlighted by the red arrows.
For quantitative evaluation, the global and local SSIM and PSNR measurements of results obtained by different synthesis loss functions and Dice scores of their corresponding segmentation results are presented in Table.~\ref{tab:synthesis_loss}, which shows that the proposed w-L2 + $\ell_{h1}$ loss function obtains higher local SSIM and Dice than the others.
\begin{figure}[t]
	\centering
	\includegraphics[width=1.0\linewidth]{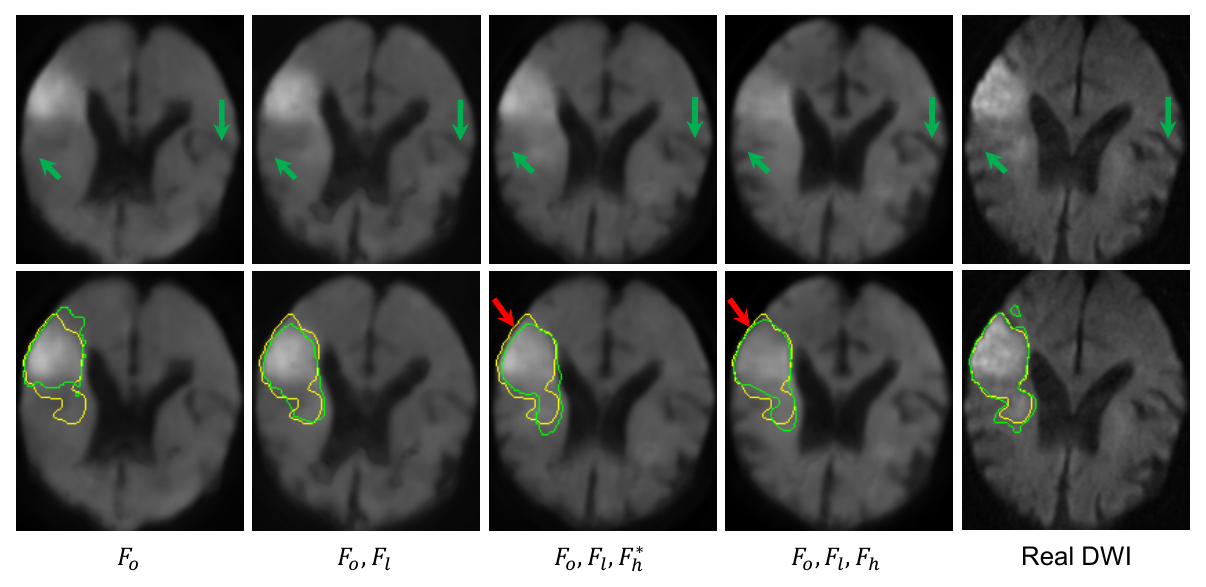}
	\caption{Visual comparison of pseudo DWI synthesized (top row) from different input images and their effect on segmentation (bottom row).  $F_o$: Concatenation of perfusion parameter maps. $F_l$: MIP of spatiotemporal CTA images. $F_h^*$ and $F_h$ are the high-level features obtained by the CNN-based feature extractor $\Phi_e$ trained without and with explicit supervision through $L_e$, respectively.  The proposed hybrid loss function  $I_g$ defined in Eq.~\ref{eq:generator_loss} was used for training. Green arrows highlight local differences of the pseudo DWI, and red arrows highlight the segmentation difference. Green and yellow curves show segmentation result and the ground truth, respectively.   }
	\label{fig:synthesis_input_visual1}
\end{figure}

\begin{table*}[t]
	\centering
	\scriptsize
	\caption{Quantitative evaluation of different inputs for pseudo DWI synthesis and their effect on segmentation. $F_o$: Concatenation of perfusion parameter maps. $F_l$: MIP of spatiotemporal CTA images. $F_h^*$ and $F_h$ are the high-level features obtained by the CNN-based feature extractor $\Phi_e$ trained without and with explicit supervision through $L_e$, respectively.
	The proposed hybrid loss function $L_g$ was used for training.}
	\begin{tabular}{l|c|c|c|c|c}
		\hline
		Input& Global SSIM & Local SSIM  &  Global PSNR & Local PSNR & Dice (\%) \\  \hline
		$F_o$ & 0.83$\pm$0.10 & 0.54$\pm$0.13 &19.26$\pm$3.32  & 13.20$\pm$3.38  & 51.25$\pm$17.43 \\
		$F_o$, $F_l$ & 0.84$\pm$0.12 & 0.56$\pm$0.15  & 20.01$\pm$3.69 & 13.90$\pm$4.04  & 53.94$\pm$14.39 \\
		$F_o$, $F_l$, $F_h^*$  & 0.84$\pm$0.11 & 0.58$\pm$0.16 & \textbf{20.16$\pm$3.97}  & 14.05$\pm$4.04 & 54.61$\pm$20.19 \\	
		$F_o$, $F_l$, $F_h$& \textbf{0.85$\pm$0.12} & \textbf{0.59$\pm$0.12} & 20.02$\pm$3.51  & \textbf{14.11$\pm$3.98} & \textbf{55.10$\pm$16.20} \\ \hline
		Real DWI &   &  &    &   & 72.17$\pm$19.54\\
		\hline
	\end{tabular}
	\label{tab:synthesis_input}
\end{table*}
\begin{figure*}[t]
	\centering
	\includegraphics[width=0.9\linewidth]{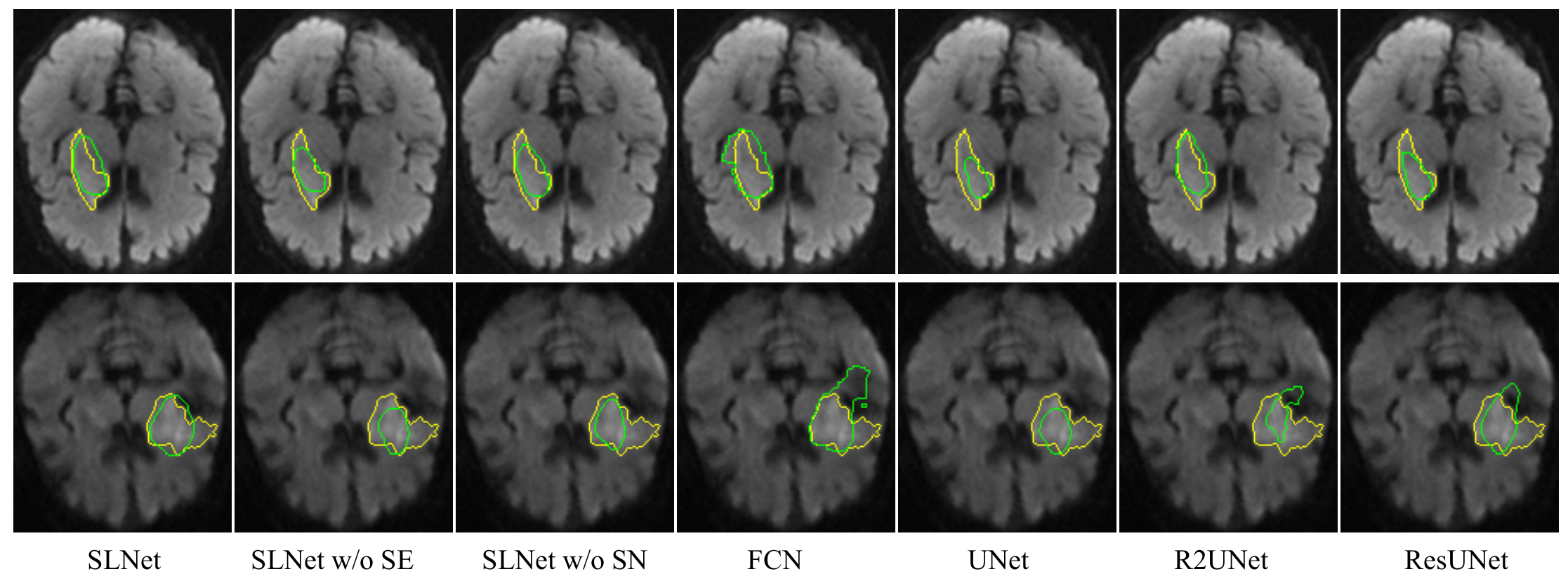}
	\caption{Visual comparison of different networks for ischemic stroke lesion segmentation. Concatenation of the CTP perfusion parameter maps ($F_o$) was used as the input of CNNs and cross entropy loss function was used for training. For better visualization, the segmentation results are shown with the real DWI images. }
	\label{fig:seg_net_visual}
\end{figure*}
\begin{table*}[t]
	\centering
	\scriptsize
	\caption{Quantitative evaluation of different networks for ischemic stroke lesion segmentation. SLNet: The proposed network for ischemic stroke lesion segmentation. Concatenation of the CTP perfusion parameter maps ($F_o$) was used as the input, and the cross entropy loss function was used for training.}
	\begin{tabular}{l|c|c|c|c|c|c}
		\hline
		Network & Parameter (M) & Precision (\%) & Recall (\%) & Dice (\%) & HD (mm) & ASSD (mm) \\  \hline
		FCN~\citep{Long2014} & 18.64 & 52.69$\pm$25.28 & 53.10$\pm$33.47& 45.75$\pm$24.59  & 48.75$\pm$30.83  & 3.78$\pm$5.37 \\
		UNet~\citep{Ronneberger2015} & 31.04 & 63.50$\pm$22.22 & 48.50$\pm$24.55 & 49.94$\pm$19.51 &21.52$\pm$13.84 &2.54$\pm$3.34\\
		R2UNet~\citep{Alom2018} &39.09 & 51.85$\pm$17.99 & 60.05$\pm$21.14 & 52.34$\pm$16.62 & 26.74$\pm$15.04 & 2.56$\pm$2.21 \\	
		ResUNet~\citep{Xiao2018} &81.91&  66.46$\pm$19.32 & 52.36$\pm$24.12  & 52.49$\pm$18.66 & 23.50$\pm$15.48 & 2.22$\pm$2.04 \\	
		SLNet & 33.84 & 51.20$\pm$22.00 & \textbf{64.20$\pm$23.99} & \textbf{54.45$\pm$21.23} & 23.97$\pm$17.61 & 2.74$\pm$3.44 \\
		SLNet (w/o SE) &31.04& \textbf{69.00$\pm$22.90} & 51.54$\pm$19.81  & 53.41$\pm$14.31  & \textbf{21.26$\pm$12.32 } & \textbf{2.16$\pm$1.90} \\
		SLNet (w/o SN) &33.84 & 68.14$\pm$21.08   & 48.32$\pm$23.31    & 52.01$\pm$20.69 & 21.50$\pm$15.63 &2.57$\pm$2.78 \\ \hline
	\end{tabular}
	\label{tab:seg_network}
\end{table*}
\subsubsection{Effect of Feature Extractor on Pseudo DWI Synthesis}
To investigate the effect of our feature extractor on the synthesized pseudo DWI, we  compared the quality of pseudo DWI images generated from different inputs: 1) the standard CTP perfusion parameter maps ($F_o$) only, i.e., without using our feature extractor; 2) concatenation of $F_o$ and our extracted low-level feature $F_l$ defined in Eq.~\ref{eq:f_l}; 3) concatenation of $F_o$, $F_l$ and $F_h^*$, where $F_h^*$ denotes the high-level feature obtained by the CNN-based feature extractor $\Phi_e$ trained without explicit supervision, i.e., $L_e$ is not used; and 4) concatenation of $F_o$, $F_l$ and $F_h$, where $F_h$ is the high-level feature obtained by $\Phi_e$ trained with explicit supervision through $L_e$. We used the proposed loss function $I_g$ (i.e., w-L2 + $\ell_{h1}$) to train the synthesis network. To additionally investigate how these synthesized results affect the segmentation, we used the standard cross entropy loss to train a UNet~\citep{Ronneberger2015} using each type of these synthesized pseudo DWI images respectively.
Fig.~\ref{fig:synthesis_input_visual1} shows a visual comparison of pseudo DWI synthesized from different input images. It can be observed that using additional $F_l$ and $F_h$ helps to improve local details of the synthesized pseudo DWI, and the result obtained by concatenation of $F_o$, $F_l$ and $F_h$ with explicit supervision lead to better image quality than the other variants, as highlighted by the green arrows.
Table~\ref{tab:synthesis_input} presents a quantitative comparison between these different inputs for pseudo DWI synthesis and the downstream segmentation, which shows that using additional low-level feature $F_l$ leads to an improvement of global and local SSIM and PSNR from using CTP perfusion parameter maps $F_o$ only. The high-level feature $F_h$ extracted by CNN and explicit supervision by $L_e$ can further lead to improved SSIM and PSNR values, which demonstrates that the proposed feature extractor making use of the raw spatiotemporal CTA images helps to obtain better synthesized pseudo DWI images.  Fig.~\ref{fig:synthesis_input_visual1} and Table~\ref{tab:synthesis_input} also show that synthesis based on $F_o$,  $F_l$ and $F_h$ leads to higher segmentation accuracy than the other variants.

\subsubsection{Comparison of Different Network for Segmentation}
To investigate the effect of network structure on our ischemic stroke lesion segmentation task, we compared our proposed SLNet with 1) SLNet w/o SE, where the SE blocks are not used in SLNet, 2) SLNet w/o SN, where the switchable normalization layers are replaced with traditional batch normalization layers in SLNet, 3) the Fully Convolutional Network (FCN)~\citep{Long2014}, 4) UNet~\citep{Ronneberger2015}, 5) Recurrent Residual UNet (R2UNet)~\citep{Alom2018}, and 6) Residual UNet (ResUNet)~\citep{Xiao2018}. We trained these networks with CTP perfusion parameter maps $F_o$ as input and used the cross entropy loss function for training. 

Fig.~\ref{fig:seg_net_visual} shows a visual comparison of segmentation results obtained by these networks, where the lesions are shown with the corresponding real DWI images for better visualization. It can be observed that it is challenging for all these networks to obtain very accurate segmentation of the ischemic stroke lesion. However, the results of our SLNet have a better overlap with the ground truth compared with the others. In the first row, the difference between different networks is relatively small. In the second row, SLNet w/o SE, SLNet w/o SN and UNet obtained more under-segmentations than SLNet, and FCN, R2UNet and ResUNet obtained more  over-segmentations than SLNet.

Quantitative comparison between these different networks is shown in Table~\ref{tab:seg_network}. The proposed SLNet achieved the highest average Dice score and Recall among all the compared networks, while SLNet w/o SE achieved slightly better HD and ASSD evaluation results.

\begin{table*}[t]
	\centering
	\scriptsize
	\caption{Quantitative evaluation of different training loss functions for ischemic stroke lesion segmentation based on our proposed SLNet. Concatenation of perfusion  parameter maps ($F_o$) was used as the input. $L_{CE}$: Cross entropy loss. $L_{WCE}$: Weighted cross entropy loss. $L_{DICE}$: Dice loss. $L_{GD}$: Generalized Dice loss. $L_{HGD}$: Hardness-aware generalized Dice loss.}
	\begin{tabular}{l|c|c|c|c|c}
		\hline
		Loss function& Precision (\%) & Recall (\%) & Dice (\%) & HD (mm) & ASSD (mm) \\  \hline
		$L_{CE}$ & 51.20$\pm$22.00 & 64.20$\pm$23.99 & 54.45$\pm$21.23 & 23.97$\pm$17.61 & 2.74$\pm$3.44  \\
		$L_{DICE}$ &\textbf{67.48$\pm$25.25} & 45.45$\pm$21.81 & 51.57$\pm$20.98 & 24.43$\pm$18.48 & 2.47$\pm$2.56 \\
		$L_{GD}$ & 55.07$\pm$22.10  & 62.13$\pm$17.09 & 54.98$\pm$17.16 & 36.87$\pm$29.61 & 3.45$\pm$3.37 \\
		$L_{HGD}$ & 52.40$\pm$20.71 & 66.83$\pm$18.98 & 55.30$\pm$17.78 & 30.86$\pm$20.10& 2.81$\pm$2.61 \\
		$L_{CE}$ + $L_{HGD}$   & 57.31$\pm$22.87  & 66.37$\pm$17.85  & 57.82$\pm$17.33& \textbf{21.58$\pm$12.69} & 1.96$\pm$1.84  \\
		$L_{WCE}$ + $L_{HGD}$ & 55.20$\pm$20.99 & \textbf{73.51$\pm$17.48}  & \textbf{59.37$\pm$15.73} & 22.29$\pm$13.67 & \textbf{1.90$\pm$2.05}  \\		\hline
	\end{tabular}
	\label{tab:seg_loss}
\end{table*}

\subsubsection{Comparison of Different Training Loss Functions for Segmentation}

We also investigate the effect of different training loss functions for the segmentation network. We refer to our proposed weighted cross entropy loss with hardness-aware generalized Dice loss as $L_{WCE} + L_{HGD}$ and compare it with 1) cross entropy loss $L_{CE}$, 2) Dice loss $L_{DICE}$~\citep{Milletari2016}, 3) generalized Dice loss $L_{GD}$~\citep{Sudre2017}, 4) hardness-weighted $L_{GD}$, which is defined in Eq.~\ref{eq:hgd} and referred to as $L_{HGD}$, and 5) a variant of the proposed loss that does not pay attention to lesion foreground (i.e., $A_i$ is 1 for every voxel), which is referred to as $L_{CE} + L_{HGD}$. We used these loss functions to train our SLNet to segment the ischemic stroke lesion from CTP perfusion parameter maps $F_o$ respectively. 

Quantitative evaluation results of these different segmentation loss functions are listed in Table~\ref{tab:seg_loss}. It can be observed that the combination of $L_{CE}$ and $L_{HGD}$ outperforms using a single loss of $L_{CE}$ or $L_{HGD}$. By enabling the network to focus more on the lesion region through $L_{WCE} + L_{HGD}$, the values of Recall and Dice are improved. Our proposed $L_{WCE} + L_{HGD}$ achieved the highest average Dice score of 59.37\%, which is a large improvement from 54.45\% achieved by the baseline of $L_{CE}$.

\subsubsection{Effect of Feature Extractor and Pseudo DWI Generator on Segmentation}

\begin{figure*}[t]
	\centering
	\includegraphics[width=0.9\linewidth]{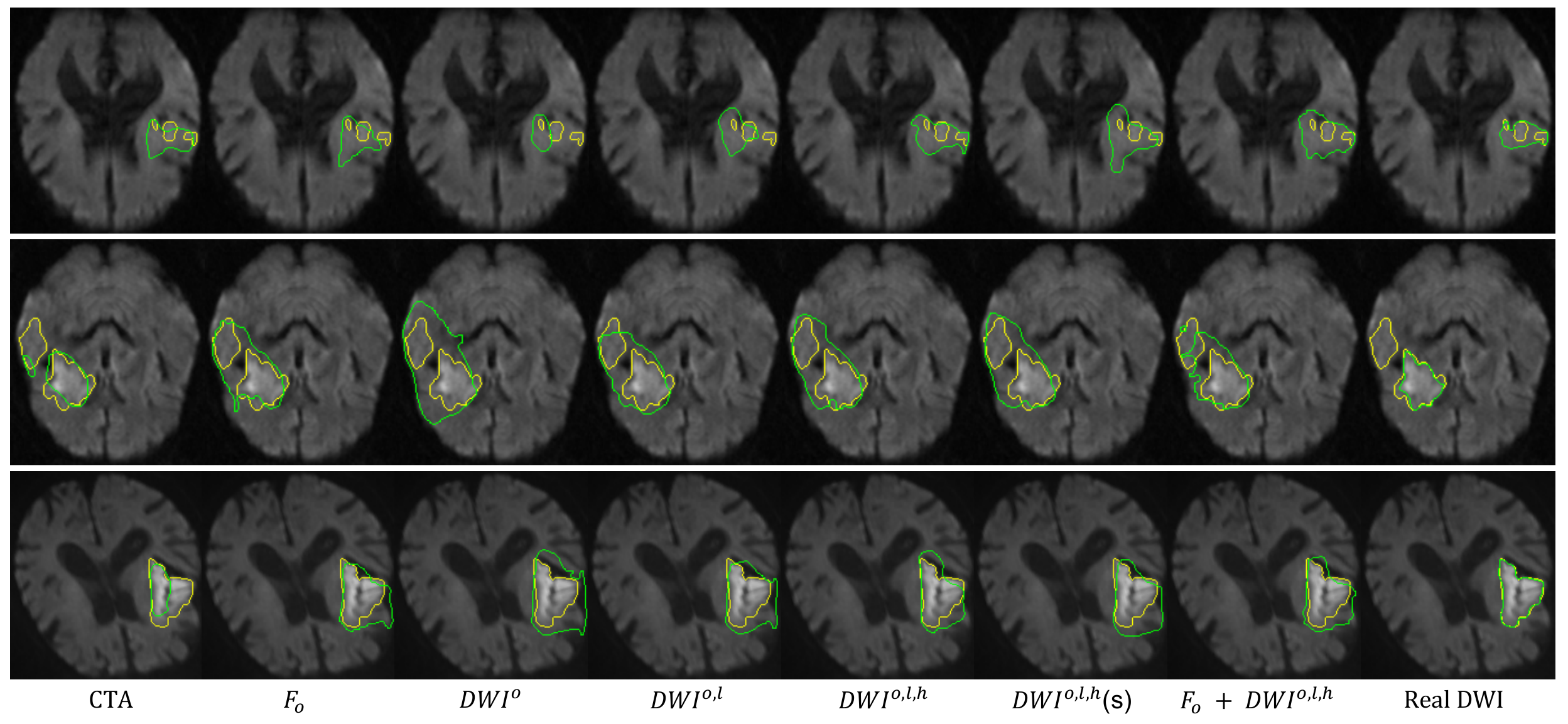}
	\caption{Visual comparison of ischemic stroke lesion segmentation from different input images. Yellow and green curves show segmentation and the ground truth, respectively. $F_o$: CTP perfusion parameter maps. DWI$^o$,  DWI$^{o, l}$ and  DWI$^{o, l, h}$ are pseudo DWI images generated from $F_o$, ($F_o$, $F_l$), and ($F_o$, $F_l$, $F_h$) respectively. DWI$^{o, l, h}$(s) is a variant of DWI$^{o, l, h}$ where our $\Phi_e$, $\Phi_g$ and $\Phi_s$ were trained subsequently rather than end-to-end. 
	For better visualization, the segmentation results are shown with the real DWI images.}
	\label{fig:seg_input_visual}
\end{figure*}
\begin{table*}[t]
	\centering
	\scriptsize
	\caption{Quantitative comparison of ischemic stroke lesion segmentation from different input images.  $F_o$:  perfusion parameter maps. DWI$^{o, l, h}$ is our proposed method with pseudo DWI synthesized from ($F_o$, $F_l$, $F_h$) as shown in Fig.~\ref{fig:framework}. DWI$^{o, l, h}$(s) is a variant of DWI$^{o, l, h}$ where our $\Phi_e$, $\Phi_g$ and $\Phi_s$ were trained subsequently rather than end-to-end. The results are based on our proposed SLNet and loss function $L_s$ defined in Eq.~\ref{eq:segment_loss}. }
	\begin{tabular}{l|c|c|c|c|c|c}
		\hline
		Input & Precision (\%) & Recall (\%) & Dice (\%) & HD (mm) & ASSD (mm) & RVE \\  \hline
		CTA &\textbf{65.63$\pm$23.08} & 58.10$\pm$18.63  & 56.10$\pm$14.22 & 25.25$\pm$16.60 & 2.45$\pm$2.26 & 0.73$\pm$0.74\\
		$F_o$ &55.20$\pm$20.99 & 73.51$\pm$17.48  & 59.37$\pm$15.73 & 22.29$\pm$13.67 & 1.90$\pm$2.05& 0.83$\pm$1.27\\
		DWI$^{o}$ & 53.68$\pm$21.27 & 74.37$\pm$16.11 & 58.32$\pm$15.74 & 19.83$\pm$13.10 &1.90$\pm$2.09 & 0.99$\pm$1.63\\
		DWI$^{o, l}$ & 58.70$\pm$22.32 & 71.04$\pm$14.75 & 60.49$\pm$16.26 & 22.79$\pm$16.90 &1.99$\pm$2.07 &0.83$\pm$1.34 \\
		DWI$^{o, l, h}$ & 61.97$\pm$21.98 & 69.52$\pm$17.89 & 62.11$\pm$17.18 & \textbf{19.27$\pm$13.17} &\textbf{1.76$\pm$2.10} &\textbf{0.68$\pm$1.36} \\
		DWI$^{o, l, h}$(s) & 57.05$\pm$20.76 & \textbf{77.80$\pm$13.56}   & \textbf{62.23$\pm$15.47} & 20.86$\pm$15.05 & 1.84$\pm$1.97 & 0.91$\pm$1.60 \\	
		$F_o$  + DWI$^{o, l, h}$& 59.06$\pm$22.25  & 71.30$\pm$15.46  & 60.54$\pm$17.17 & 22.42$\pm$19.58 & 2.07$\pm$2.95 & 0.83$\pm$1.33 \\ \hline	
	    Real DWI & 85.07$\pm$19.35  & 77.34$\pm$15.03  & 79.72$\pm$15.53 & 15.90$\pm$14.13 & 1.35$\pm$2.77 & 0.24$\pm$0.24\\\hline
	\end{tabular}
	\label{tab:seg_input}
\end{table*}
With our proposed feature extraction and image synthesis method, we evaluate the value of our pseudo DWI generated from $F_o$, $F_l$ and $F_h$ for ischemic stroke lesion segmentation, where the pseudo DWI is referred to as DWI$^{o,l,h}$. We compared segmentation from  DWI$^{o,l,h}$ with segmentation from 1) raw CTA images that were temporally cropped and down-sampled (i.e., $I^*$ as described in Section~\ref{extractor}), 2) CTP perfusion  parameter maps $F_o$, 3) DWI$^o$ that refers to pseudo DWI generated from $F_o$, 4) DWI$^{o,l}$ that refers to pseudo DWI generated from $F_o$ and $F_l$, and 5) concatenation of $F_o$ and DWI$^{o,l,h}$. We used these different setting of synthesized pseudo DWI images for end-to-end training respectively, where the overall loss function in Eq.~\ref{eq:overall_loss} combined with our SLNet was used for segmentation. We also compared DWI$^{o,l,h}$ with its variant DWI$^{o,l,h}$(s) that refers to our $\Phi_e$, $\Phi_g$ and $\Phi_s$ were trained subsequently rather than end-to-end. Additionally, we trained SLNet with real DWI images to investigate the gap between segmentation from synthesized pseudo DWI images and from real DWI images.

Fig.~\ref{fig:seg_input_visual}
presents a visual comparison between ischemic stroke lesion segmentation results from different input images, which shows that the results segmented from our synthesized pseudo DWI images are better than those of other variants.
Table~\ref{tab:seg_input} presents the quantitative evaluation results. It shows that using DWI$^o$ generated from CTP  perfusion parameter maps leads to a slightly decreased segmentation accuracy. By using additional features $F_l$ and $F_h$ extracted from the raw spatiotemporal CTA images for synthesis, DWI$^{o,l}$  and  DWI$^{o,l,h}$ lead to an improvement of Dice score respectively.  Table~\ref{tab:seg_input} shows that using DWI$^{o,l,h}$ outperformed the other variants. 
The average Dice scores for segmentation from original CTA images, perfusion parameter maps (i.e., $F_o$), synthesized pseudo DWI based on our proposed method (i.e., DWI$^{o,l,h}$) and real DWI are 56.10\%, 59.37\%, 62.11\% and 79.72\%, respectively. The corresponding Hausdorff Distance values are 25.25~mm, 22.29~mm, 19.27~mm and 15.90~mm, respectively.  We found that adding $F_o$ to DWI$^{o, h, l}$ leads to a reduced segmentation performance compared with using DWI$^{o, h, l}$ only. This is due to that using $F_o$ performs worse than using DWI$^{o, h, l}$, and a combination of them just obtains a segmentation accuracy above that of using $F_o$ and below that of using  DWI$^{o, h, l}$. It can be observed from Table~\ref{tab:seg_input} that  DWI$^{o,l,h}$ and DWI$^{o,l,h}$(s) obtained very close segmentation accuracy in terms of Dice. However, DWI$^{o,l,h}$ achieved smaller HD and ASSD values than DWI$^{o,l,h}$(s). 
\begin{figure*}
	\centering
	\subfloat[Small lesions ($<$ 10 CC) \label{fig:1a}]
	{\includegraphics[width=0.3\textwidth]{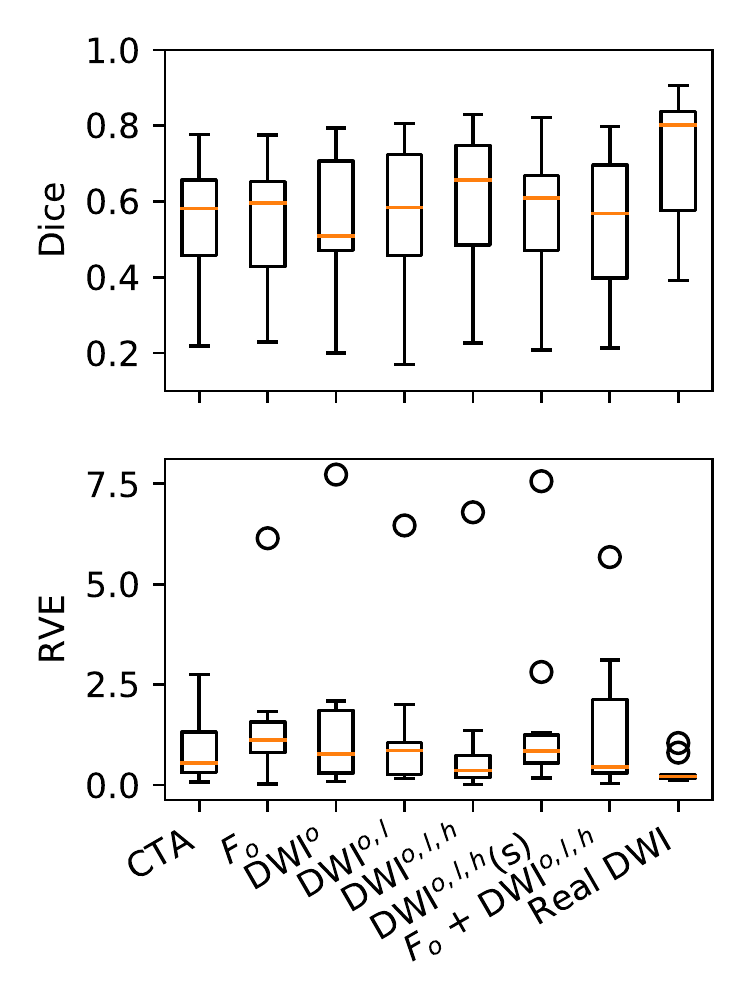}}
	\hfill
	\subfloat[Medium lesions (10 to 50 CC) \label{fig:1b}] {\includegraphics[width=0.3\textwidth]{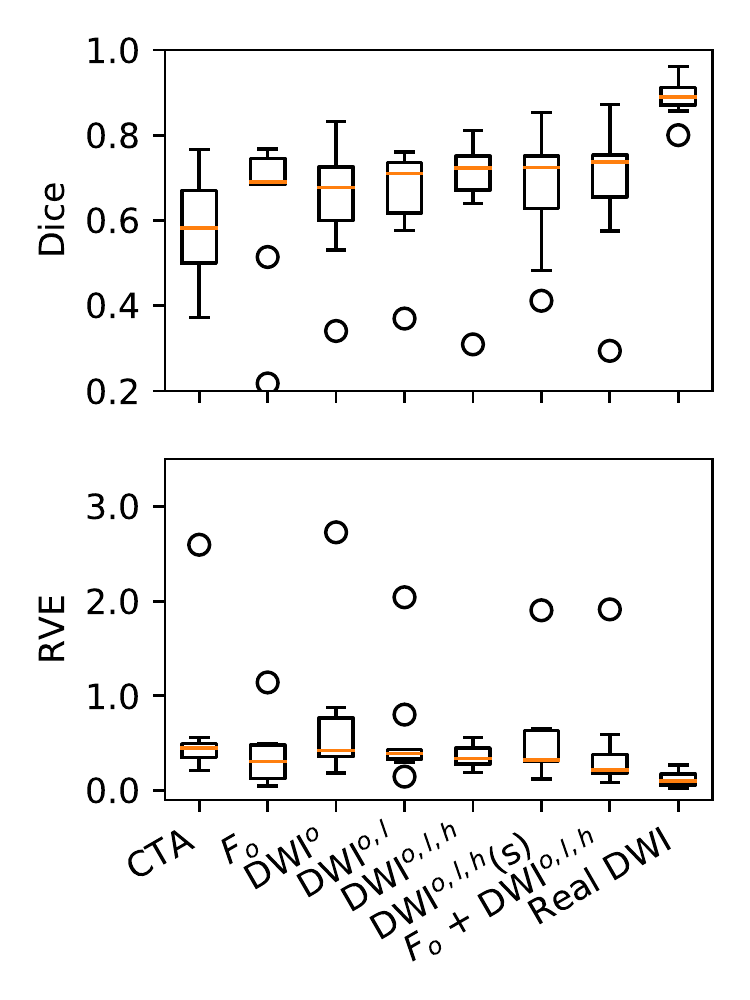}}
	\hfill
	\subfloat[Large lesions ($>$ 50 CC) \label{fig:1c}]
	{\includegraphics[width=0.3\textwidth]{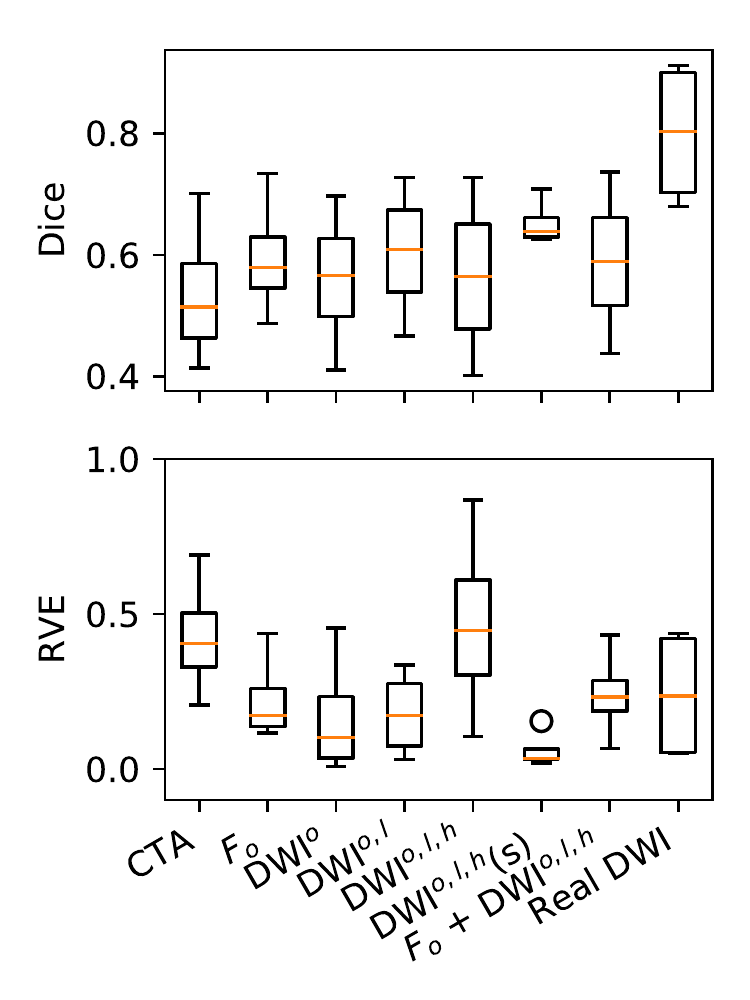}}
	\caption{Dice and RVE for lesions at three scales segmented from different types of images.  $F_o$: perfusion parameter maps. DWI$^{o, l, h}$ is our proposed method with pseudo DWI synthesized from ($F_o$, $F_l$, $F_h$) as shown in Fig.~\ref{fig:framework}. DWI$^{o, l, h}$(s) is a variant of DWI$^{o, l, h}$ where our $\Phi_e$, $\Phi_g$ and $\Phi_s$ were trained subsequently rather than end-to-end. The results are based on our proposed SLNet and loss function $L_s$. } 
	\label{fig:boxplot_dice}
\end{figure*} 

As the ischemic stroke lesions vary largely in sizes, we investigated the segmentation performance at different lesion scales. We divided the local testing set into three groups: 1) 9 images with small lesions ($<$ 10~CC), 2) 10 images with medium lesions (10 - 50~CC) and 3) 4 images with large lesions ($>$ 50~CC). For evaluation, we additionally measured the Relative Volume Error (RVE): $RVE = abs(V_g - V_s) / V_g$, where $V_g$ and $V_s$ are the volume of a ground truth lesion and the segmented lesion, respectively. Table~\ref{tab:seg_input} shows that DWI$^{o,l,h}$ obtained a lower average RVE value than the others except for the real DWI.
Fig.~\ref{fig:boxplot_dice} shows the distributions of Dice and RVE in these three groups.  The average Dice values achieved by our proposed method (i.e., DWI$^{o, l, h}$) for these three groups were 59.50\%, 68.87\% and 56.44\% respectively. The lower performance in the small and large groups indicate that it remains difficult for the proposed method to deal with extreme cases with small and very large lesions.

\subsection{Comparison with Other ISLES Participants} 

\begin{table}[t]
	\centering
	\small
	\caption{Quantitative comparison of the top five methods for ISLES 2018 testing set. }
	\begin{tabular}{l|c|c|c}
		\hline
		Method & Dice & Precision & Recall \\  \hline
		Ours & 0.51 $\pm$ 0.31 & 0.55 $\pm$ 0.36  & 0.55 $\pm$ 0.34 \\
		\cite{Liu2018a} & 0.49 $\pm$ 0.31 & 0.56 $\pm$ 0.37  & 0.53 $\pm$ 0.33 \\
		Chen et al. & 0.48 $\pm$ 0.32 & 0.59 $\pm$ 0.38  & 0.46 $\pm$ 0.33 \\	
		Hu et al. & 0.47 $\pm$ 0.31 & 0.56 $\pm$ 0.37  & 0.47 $\pm$ 0.33 \\
		Garcia et al. & 0.47 $\pm$ 0.31 & 0.56 $\pm$ 0.37  & 0.47 $\pm$ 0.33 \\
		\hline
	\end{tabular}
	\label{tab:compare_with_others}
\end{table}
We also trained our proposed method with the entire ISLES 2018 training set, and submitted the segmentation results of ISLES 2018 testing set to the online evaluation platform for quantitative evaluation. According to the  ISLES 2018 leaderboard\footnote{\url{https://www.smir.ch/ISLES/Start2018}}, our method achieved the top performance among 62 teams. Table~\ref{tab:compare_with_others} lists the quantitative evaluation results of the top five methods\footnote{Listed in the 'Results' section of \url{http://www.isles-challenge.org}} for  ISLES 2018, where our method outperformed the others with an average Dice score of 0.51. \cite{Liu2018a} also used a CNN to generate pseudo DWI for segmentation, but only from CTP perfusion parameter maps with GAN, and the achieved Dice and Recall are lower than ours. The other three methods segmented the ischemic stroke lesion from CTP perfusion parameter maps directly. Chen et al.\footnote{\url{http://www.isles-challenge.org/articles/Yu_Chen.pdf}} used an ensemble of multiple networks combined with several data augmentation methods. Hu et al.\footnote{\url{http://www.isles-challenge.org/articles/Xiaojun_Hu.pdf}} proposed a multi-level 3D refinement module trained with curriculum learning.
Clerigues et al.\footnote{\url{http://www.isles-challenge.org/articles/albert.pdf}} also used an ensemble of multiple networks, and employed a patch sampling strategy to alleviate class imbalance.

\section{Discussion and Conclusion}\label{sec:discussion}
Due to the low contrast and low resolution of CTP perfusion parameter maps, it is challenging to directly use these images for ischemic stroke lesion segmentation. Transferring  the perfusion parameter maps to pseudo DWI images via image synthesis is a promising way for the segmentation task, as DWI images have a better contrast between the lesion and the background and they are used for obtaining the ground truth ischemic stroke lesion region. The ISLES 2018 finalist and our experiments showed that pseudo DWI-based segmentation methods outperformed direct segmentation from perfusion parameter maps. 

The quality of the synthesized pseudo DWI images has a large impact on the segmentation performance. A good contrast with enhanced and preserved lesion information in the pseudo DWI is important for good segmentation results. Though deep learning for image synthesis has achieved very good performance in other tasks~\citep{Frangi2018}, the synthesis of pseudo DWI with ischemic stroke lesion in this study is still challenging due to the low quality of perfusion parameter maps and a small number of training images. To alleviate this problem, we used two strategies. First, we exploited information in the raw spatiotemporal CTA images by extracting low-level and high-level features in additional to the perfusion parameter maps. Results show that this helps to obtain higher pseudo DWI quality and higher segmentation accuracy than using perfusion parameter maps only, as demonstrated in Table~\ref{tab:synthesis_input} and Table~\ref{tab:seg_input}. From Fig.~\ref{fig:synthesis_input_visual1} and Table~\ref{tab:synthesis_input}, we find that using an explicit supervision on the feature extractor leads to some improvement of segmentation accuracy, but the difference was not significant. This phenomenon is expected as the explicit supervision serves as a deep supervision. When it is not used, the feature extractor can also be updated based on the loss function, and the deep supervision mainly helps to improve the convergence during training.
Second, we designed a weighted loss function that pays attention to the lesion region so that the quality of the generated lesion is highlighted. It is  combined with a high-level contextual loss function that encourages global and high-level consistency between the generated pseudo DWI and the ground truth DWI. Results in Table~\ref{tab:synthesis_loss} show that this leads to an improvement of local SSIM around the lesion region. However, 
we found that our synthesized pseudo DWI images are still not as good as the real DWI images. For example, Table~\ref{tab:synthesis_loss} and Table~\ref{tab:synthesis_input} indicate that the PSNR numbers are not very high. This is mainly due to that the high-frequency components in the real DWI images are not well synthesized, as shown in Fig.~\ref{fig:synthesis_loss_visual1} and Fig.~\ref{fig:synthesis_input_visual1}. The high-frequency components are related to local fine-grained details, noises and some artifacts. As demonstrated by~\cite{Xu2019}, CNNs capture low-frequency components at the early stage of training, and then capture high-frequency components and tend to overfit at the late stage of training. During the training with our relatively small dataset, we used the best performing checkpoint on the validation set for testing to minimize the risk of under-fitting or over-fitting. As an incidental effect, we found that the synthesized pseudo DWI images related to that checkpoint did not have many high-frequency components.
It is of interest to further improve the pseudo DWI quality, which has a promising to obtain better segmentation results. As the synthesized pseudo DWI and real DWI can be regarded as coming from two different domains, some domain adaptation methods~\citep{Perone2019} can be used in the future to obtain better segmentation performance with pseudo DWI.

For segmentation networks, by using switchable normalization and SE block based on channel attention, the segmentation Dice and Recall are improved with a marginal increase of parameter number, as shown in Table~\ref{tab:seg_network}. The loss function for training the segmentation network also has a large impact on the segmentation performance. Our weighted cross entropy loss function $L_{WCE}$ pays more attention to the lesion region and helps to alleviate the imbalance between the foreground and the background. The hardness-aware generalized Dice loss $L_{HGD}$ automatically gives higher weights to harder samples. A combination of $L_{WCE}$ and $L_{HGD}$ considers pixel-wise and region-level accuracy simultaneously, which leads to better Dice, Recall and ASSD than the other variants as shown in Table~\ref{tab:seg_loss}. It should be noticed that the Hausdorff distance of our results is still high. To address this problem, using  Hausdorff distance-based loss functions~\citep{Kervadec2019} or high-level constraints~\citep{Oktay2017} are potential solutions.

Our high-level feature extraction, pseudo DWI  generation and lesion segmentation modules are trained end-to-end so that they are updated simultaneously and adaptive to each other with a high coherence. This makes the training process more efficient than training these modules subsequently. Results in Fig.~\ref{fig:seg_input_visual} and Table~\ref{tab:seg_input} show that the end-to-end training also benefits the final segmentation performance. However, a drawback of end-to-end training is that these modules become less portable as a change of the segmentation network requires the whole system to be trained again. Subsequent training would make the system more modular and is preferred in a scenario where there is a high demand for replacing some of these modules. For example, the segmentation network can be replaced when more training images become available without retraining the feature extractor and pseudo DWI generator. In this paper, as the training set had a small size and was fixed during the study, we chose the end-to-end training strategy due to its efficiency and better segmentation performance.

Comparing Table~\ref{tab:seg_input} and Table~\ref{tab:compare_with_others}, we observe that there is a performance drop between our local testing set and the official testing set of ISLES 2018. This indicates some overfitting of the proposed method. The overfitting could be attributed to a couple of reasons. First, the training set was relatively small and each image only contained 5.34 slices in average. Second, our method relies on image synthesis as an intermediate step, and there might be a domain shift between synthesized pseudo DWI images and real DWI images. The two steps of synthesis and segmentation are prone to accumulate the prediction error and possibility of overfitting. To deal with this problem, using some advanced data augmentation methods~\citep{Abdulkadir2016,Frid-Adar2018} and additional regularizations such as auxiliary tasks~\citep{Myronenko2018} and volume constraints~\citep{Kervadec2019a} could be potential approaches. Fig.~\ref{fig:boxplot_dice} shows that the proposed method did not segment well on large lesions, which is mainly because the large lesion group contained only few cases (i.e., 4 images for testing), and it was not statistically significant to evaluate the segmentation performance for that group. In the future, a larger dataset could be used for a better evaluation.

In conclusion, to deal with the problem of ischemic stroke lesion segmentation from CTP images, we propose a novel framework using synthesized pseudo DWI images for better segmentation results. We propose a feature extractor that obtains both a low-level and a high-level compact representation of the raw spatiotemporal CTA images, and combine them with the CTP perfusion parameter maps for better pseudo DWI synthesis quality. We also propose to pay more attention to the lesion region and encourage high-level similarity for synthesis of pseudo DWI with stroke lesions. A network with switchable normalization and channel calibration trained with hardness-aware generalized Dice loss is proposed for the final segmentation from synthesized pseudo DWI. Extensive experimental results on ISLES 2018  dataset  showed that our method using synthesized pseudo DWI outperformed methods using CTA images or perfusion parameter maps directly for ischemic stroke lesion segmentation, and demonstrated that our feature extractor helps to obtain better synthesized pseudo DWI quality that leads to higher  segmentation accuracy. The proposed automatic segmentation framework has a potential for improving diagnosis and treatment of the ischemic stroke in a timely fashion, especially in acute units with limited availability of DWI scanning.

\section{Acknowledgements}\label{sec:acknowledgements}
This work was supported by the National Natural Science Foundation of China funding [81771921, 61901084].

\bibliographystyle{model2-names}
\bibliography{./reference/brain_lesion_seg}







\end{document}